\documentclass[conference]{IEEEtran}
\usepackage {amssymb,rotating,graphicx,cite,subeqn, calc,color,multirow}
\newtheorem{example}{\bf Example}

\newcommand\JSAC{{\em IEEE Journal on Selected Areas in Communications,} vol. }

\begin{document}

\title{Interference Cancelation in Non-coherent CDMA Systems Using Parallel
Iterative Algorithms}

\author{\IEEEauthorblockN{Kamal Shahtalebi}
\IEEEauthorblockA{Department of Information Technology\\
The University of Isfahan, Isfahan\\ Iran, Postal Code:
81746-73441 \\
Email: shahtalebi@eng.ui.ac.ir} \and \IEEEauthorblockN{Hamidreza
Saligheh Rad}
\IEEEauthorblockA{School of Engineering and Applied\\
Sciences, Harvard University\\ Cambridge MA, 02138, USA\\
Email: hamid@seas.harvard.edu} \and \IEEEauthorblockN{Gholam Reza
Bakhshi} \IEEEauthorblockA{Department of Electrical and Computer\\
Engineering, Yazd University, Yazd, Iran\\
Email: farbakhshi@yahoo.com}}

\maketitle

\begin{abstract}
Parallel least mean square-partial parallel interference cancelation
(PLMS-PPIC) is a partial interference cancelation which employs
adaptive multistage structure \cite{cohpaper}. In this algorithm the
channel phases for all users are assumed to be known. Having only
their quarters in $(0,2\pi)$, a modified version of PLMS-PPIC is
proposed in this paper to simultaneously estimate the channel phases
and the cancelation weights. Simulation examples are given in the
cases of balanced, unbalanced and time varying channels to show the
performance of the modified PLMS-PPIC method.
\end{abstract}
%\clearpage \listoftables\listoffigures
\section{Introduction}\label{S1}

The multiple access interferences (MAI) is the root of user
limitation in CDMA systems \cite{R1,R3}. The parallel least mean
square-partial parallel interference cancelation (PLMS-PPIC) method
is a multiuser detector for code division multiple access (CDMA)
receivers which reduces the effect of MAI in bit detection. In this
method and similar to its former versions like LMS-PPIC \cite{R5}
(see also \cite{RR5}), a weighted value of the MAI of other users is
subtracted before making the decision for a specific user in
different stages \cite{cohpaper}. In both of these methods, the
normalized least mean square (NLMS) algorithm is engaged
\cite{Haykin96}. The $m^{\rm th}$ element of the weight vector in
each stage is the true transmitted binary value of the $m^{\rm th}$
user divided by its hard estimate value from the previous stage. The
magnitude of all weight elements in all stages are equal to unity.
Unlike the LMS-PPIC, the PLMS-PPIC method tries to keep this
property in each iteration by using a set of NLMS algorithms with
different step-sizes instead of one NLMS algorithm used in LMS-PPIC.
In each iteration, the parameter estimate of the NLMS algorithm is
chosen whose element magnitudes of cancelation weight estimate have
the best match with unity. In PLMS-PPIC implementation it is assumed
that the receiver knows the phases of all user channels. However in
practice, these phases are not known and should be estimated. In
this paper we improve the PLMS-PPIC procedure \cite{cohpaper} in
such a way that when there is only a partial information of the
channel phases, this modified version simultaneously estimates the
phases and the cancelation weights. The partial information is the
quarter of each channel phase in $(0,2\pi)$.

The rest of the paper is organized as follows: In section \ref{S4}
the modified version of PLMS-PPIC with capability of channel phase
estimation is introduced. In section \ref{S5} some simulation
examples illustrate the results of the proposed method. Finally the
paper is concluded in section \ref{S6}.

\section{Multistage Parallel Interference Cancelation: Modified PLMS-PPIC Method}\label{S4}

We assume $M$ users synchronously send their symbols
$\alpha_1,\alpha_2,\cdots,\alpha_M$ via a base-band CDMA
transmission system where $\alpha_m\in\{-1,1\}$. The $m^{th}$ user
has its own code $p_m(.)$ of length $N$, where $p_m(n)\in \{-1,1\}$,
for all $n$. It means that for each symbol $N$ bits are transmitted
by each user and the processing gain is equal to $N$. At the
receiver we assume that perfect power control scheme is applied.
Without loss of generality, we also assume that the power gains of
all channels are equal to unity and users' channels do not change
during each symbol transmission (it can change from one symbol
transmission to the next one) and the channel phase $\phi_m$ of
$m^{th}$ user is unknown for all $m=1,2,\cdots,M$ (see
\cite{cohpaper} for coherent transmission). According to the above
assumptions the received signal is
\begin{equation}
\label{e1} r(n)=\sum\limits_{m=1}^{M}\alpha_m
e^{j\phi_m}p_m(n)+v(n),~~~~n=1,2,\cdots,N,
\end{equation}
where $v(n)$ is the additive white Gaussian noise with zero mean and
variance $\sigma^2$. Multistage parallel interference cancelation
method uses $\alpha^{s-1}_1,\alpha^{s-1}_2,\cdots,\alpha^{s-1}_M$,
the bit estimates outputs of the previous stage, $s-1$, to estimate
the related MAI of each user. It then subtracts it from the received
signal $r(n)$ and makes a new decision on each user variable
individually to make a new variable set
$\alpha^{s}_1,\alpha^{s}_2,\cdots,\alpha^{s}_M$ for the current
stage $s$. Usually the variable set of the first stage (stage $0$)
is the output of a conventional detector. The output of the last
stage is considered as the final estimate of transmitted bits. In
the following we explain the structure of a modified version of the
PLMS-PIC method \cite{cohpaper} with simultaneous capability of
estimating the cancelation weights and the channel phases.

Assume $\alpha_m^{(s-1)}\in\{-1,1\}$ is a given estimate of
$\alpha_m$ from stage $s-1$. Define
\begin{equation}
\label{e6} w^s_{m}=\frac{\alpha_m}{\alpha_m^{(s-1)}}e^{j\phi_m}.
\end{equation}
From (\ref{e1}) and (\ref{e6}) we have
\begin{equation}
\label{e7} r(n)=\sum\limits_{m=1}^{M}w^s_m\alpha^{(s-1)}_m
p_m(n)+v(n).
\end{equation}
Define
\begin{subequations}
\begin{eqnarray}
\label{e8} W^s&=&[w^s_{1},w^s_{2},\cdots,w^s_{M}]^T,\\
\label{e9}
\!\!\!\!\!\!\!\!\!\!\!\!\!\!\!\!\!\!\!\!X^{s}(n)\!\!\!&=&\!\!\![\alpha^{(s-1)}_1p_1(n),\alpha^{(s-1)}_2p_2(n),\cdots,\alpha^{(s-1)}_Mp_M(n)]^T.
\end{eqnarray}
\end{subequations}
where $T$ stands for transposition. From equations (\ref{e7}),
(\ref{e8}) and (\ref{e9}), we have
\begin{equation}
\label{e10} r(n)=W^{s^T}X^{s}(n)+v(n).
\end{equation}
Given the observations $\{r(n),X^{s}(n)\}^{N}_{n=1}$, in modified
PLMS-PPIC, like the PLMS-PPIC \cite{cohpaper}, a set of NLMS
adaptive algorithm are used to compute
\begin{equation}
\label{te1} W^{s}(N)=[w^{s}_1(N),w^{s}_2(N),\cdots,w^{s}_M(N)]^T,
\end{equation}
which is an estimate of $W^s$ after iteration $N$. To do so, from
(\ref{e6}), we have
\begin{equation}
\label{e13} |w^s_{m}|=1 ~~~m=1,2,\cdots,M,
\end{equation}
which is equivalent to
\begin{equation}
\label{e14} \sum\limits_{m=1}^{M}||w^s_{m}|-1|=0.
\end{equation}
We divide $\Psi=\left(0,1-\sqrt{\frac{M-1}{M}}\right]$, a sharp
range for $\mu$ (the step-size of the NLMS algorithm) given in
\cite{sg2005}, into $L$ subintervals and consider $L$ individual
step-sizes $\Theta=\{\mu_1,\mu_2,\cdots,\mu_L\}$, where
$\mu_1=\frac{1-\sqrt{\frac{M-1}{M}}}{L}, \mu_2=2\mu_1,\cdots$, and
$\mu_L=L\mu_1$. In each stage, $L$ individual NLMS algorithms are
executed ($\mu_l$ is the step-size of the $l^{th}$ algorithm). In
stage $s$ and at iteration $n$, if
$W^{s}_k(n)=[w^s_{1,k},\cdots,w^s_{M,k}]^T$, the parameter estimate
of the $k^{\rm th}$ algorithm, minimizes our criteria, then it is
considered as the parameter estimate at time iteration $n$. In other
words if the next equation holds
\begin{equation}
\label{e17} W^s_k(n)=\arg\min\limits_{W^s_l(n)\in I_{W^s}
}\left\{\sum\limits_{m=1}^{M}||w^s_{m,l}(n)|-1|\right\},
\end{equation}
where $W^{s}_l(n)=W^{s}(n-1)+\mu_l \frac{X^s(n)}{\|X^s(n)\|^2}e(n),
~~~ l=1,2,\cdots,k,\cdots,L-1,L$ and
$I_{W^s}=\{W^s_1(n),\cdots,W^s_L(n)\}$, then we have
$W^s(n)=W^s_k(n)$, and therefore all other algorithms replace their
weight estimate by $W^{s}_k(n)$. At time instant $n=N$, this
procedure gives $W^s(N)$, the final estimate of $W^s$, as the true
parameter of stage $s$.

Now consider $R=(0,2\pi)$ and divide it into four equal parts
$R_1=(0,\frac{\pi}{2})$, $R_2=(\frac{\pi}{2},\pi)$,
$R_3=(\pi,\frac{3\pi}{2})$ and $R_4=(\frac{3\pi}{2},2\pi)$. The
partial information of channel phases (given by the receiver) is in
a way that it shows each $\phi_m$ ($m=1,2,\cdots,M$) belongs to
which one of the four quarters $R_i,~i=1,2,3,4$. Assume
$W^{s}(N)=[w^{s}_1(N),w^{s}_2(N),\cdots,w^{s}_M(N)]^T$ is the weight
estimate of the modified algorithm PLMS-PPIC at time instant $N$ of
the stage $s$. From equation (\ref{e6}) we have
\begin{equation}
\label{tt3}
\phi_m=\angle({\frac{\alpha^{(s-1)}_m}{\alpha_m}w^s_m}).
\end{equation}
We estimate $\phi_m$ by $\hat{\phi}^s_m$, where
\begin{equation}
\label{ee3}
\hat{\phi}^s_m=\angle{(\frac{\alpha^{(s-1)}_m}{\alpha_m}w^s_m(N))}.
\end{equation}
Because $\frac{\alpha^{(s-1)}_m}{\alpha_m}=1$ or $-1$, we have
\begin{eqnarray}
\hat{\phi}^s_m=\left\{\begin{array}{ll} \angle{w^s_m(N)} &
\mbox{if}~
\frac{\alpha^{(s-1)}_m}{\alpha_m}=1\\
\pm\pi+\angle{w^s_m(N)} & \mbox{if}~
\frac{\alpha^{(s-1)}_m}{\alpha_m}=-1\end{array}\right.
\end{eqnarray}
Hence $\hat{\phi}^s_m\in P^s=\{\angle{w^s_m(N)},
\angle{w^s_m(N)+\pi, \angle{w^s_m(N)}-\pi}\}$. If $w^s_m(N)$
sufficiently converges to its true value $w^s_m$, the same region
for $\hat{\phi}^s_m$ and $\phi_m$ is expected. In this case only one
of the three members of $P^s$ has the same region as $\phi_m$. For
example if $\phi_m \in (0,\frac{\pi}{2})$, then $\hat{\phi}^s_m \in
(0,\frac{\pi}{2})$ and therefore only $\angle{w^s_m(N)}$ or
$\angle{w^s_m(N)}+\pi$ or $\angle{w^s_m(N)}-\pi$ belongs to
$(0,\frac{\pi}{2})$. If, for example, $\angle{w^s_m(N)}+\pi$ is such
a member between all three members of $P^s$, it is the best
candidate for phase estimation. In other words,
\[\phi_m\approx\hat{\phi}^s_m=\angle{w^s_m(N)}+\pi.\]
We admit that when there is a member of $P^s$ in the quarter of
$\phi_m$, then $w^s_m(N)$ converges. What would happen when non of
the members of $P^s$ has the same quarter as $\phi_m$? This
situation will happen when the absolute difference between $\angle
w^s_m(N)$ and $\phi_m$ is greater than $\pi$. It means that
$w^s_m(N)$ has not converged yet. In this case where we can not
count on $w^s_m(N)$, the expected value is the optimum choice for
the channel phase estimation, e.g. if $\phi_m \in (0,\frac{\pi}{2})$
then $\frac{\pi}{4}$ is the estimation of the channel phase
$\phi_m$, or if $\phi_m \in (\frac{\pi}{2},\pi)$ then
$\frac{3\pi}{4}$ is the estimation of the channel phase $\phi_m$.
The results of the above discussion are summarized in the next
equation
\begin{eqnarray}
\nonumber \hat{\phi}^s_m = \left\{\begin{array}{llll} \angle
{w^s_m(N)} & \mbox{if}~
\angle{w^s_m(N)}, \phi_m\in R_i,~~i=1,2,3,4\\
\angle{w^s_m(N)}+\pi & \mbox{if}~ \angle{w^s_m(N)}+\pi, \phi_m\in
R_i,~~i=1,2,3,4\\
\angle{w^n_m(N)}-\pi & \mbox{if}~ \angle{w^s_m(N)}-\pi, \phi_m\in
R_i,~~i=1,2,3,4\\
\frac{(i-1)\pi+i\pi}{4} & \mbox{if}~ \phi_m\in
R_i,~~\angle{w^s_m(N)},\angle
{w^s_m(N)}\pm\pi\notin R_i,~~i=1,2,3,4.\\
\end{array}\right.
\end{eqnarray}
Having an estimation of the channel phases, the rest of the proposed
method is given by estimating $\alpha^{s}_m$ as follows:
\begin{equation}
\label{tt4}
\alpha^{s}_m=\mbox{sign}\left\{\mbox{real}\left\{\sum\limits_{n=1}^{N}
q^s_m(n)e^{-j\hat{\phi}^s_m}p_m(n)\right\}\right\},
\end{equation}
where
\begin{equation} \label{tt5}
q^{s}_{m}(n)=r(n)-\sum\limits_{m^{'}=1,m^{'}\ne
m}^{M}w^{s}_{m^{'}}(N)\alpha^{(s-1)}_{m^{'}} p_{m^{'}}(n).
\end{equation}
The inputs of the first stage $\{\alpha^{0}_m\}_{m=1}^M$ (needed for
computing $X^1(n)$) are given by
\begin{equation}
\label{qte5}
\alpha^{0}_m=\mbox{sign}\left\{\mbox{real}\left\{\sum\limits_{n=1}^{N}
r(n)e^{-j\hat{\phi}^0_m}p_m(n)\right\}\right\}.
\end{equation}
Assuming $\phi_m\in R_i$, then
\begin{equation}
\label{qqpp} \hat{\phi}^0_m =\frac{(i-1)\pi+i\pi}{4}.
\end{equation}
Table \ref{tab4} shows the structure of the modified PLMS-PPIC
method. It is to be notified that
\begin{itemize}
\item Equation (\ref{qte5}) shows the conventional bit detection
method when the receiver only knows the quarter of channel phase in
$(0,2\pi)$. \item With $L=1$ (i.e. only one NLMS algorithm), the
modified PLMS-PPIC can be thought as a modified version of the
LMS-PPIC method.
\end{itemize}

In the following section some examples are given to illustrate the
effectiveness of the proposed method.

\section{Simulations}\label{S5}

In this section we have considered some simulation examples.
Examples \ref{ex2}-\ref{ex4} compare the conventional, the modified
LMS-PPIC and the modified PLMS-PPIC methods in three cases: balanced
channels, unbalanced channels and time varying channels. In all
examples, the receivers have only the quarter of each channel phase.
Example \ref{ex2} is given to compare the modified LMS-PPIC and the
PLMS-PPIC in the case of balanced channels.

\begin{figure} \centering
\includegraphics[width=.45\textwidth]{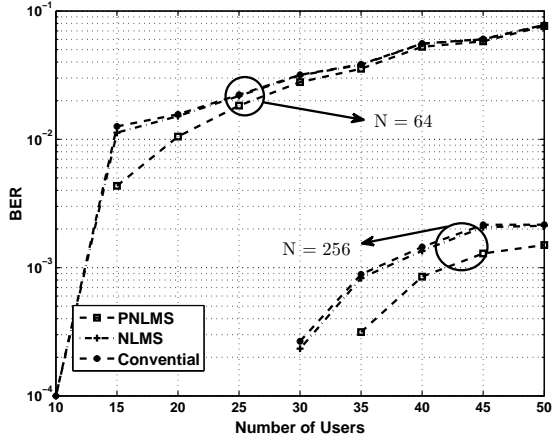}
\caption{The BER of the conventional, the modified LMS-PPIC and the
modified PLMS-PPIC methods versus the system load in balanced
channel, using two stages for $N=64$ and
$N=256$.}\label{Figexp1NonCoh}
\end{figure}

\begin{example}{\it Balanced channels}:
\label{ex2}
\begin{table}
\caption{Channel phase estimate of the first user (example
\ref{ex2})} \label{tabex5} \centerline{{
\begin{tabular}{|c|c|c|c|c|}
\hline
\multirow{6}{*}{\rotatebox{90}{$\phi_m=\frac{3\pi}{8},M=15~~$}} & N(Iteration) & Stage Number& NLMS & PNLMS  \\
&&&&\\
\cline{2-5} & \multirow{2}{*}{64}& s = 2 &  $\hat{\phi}^s_m=\frac{3.24\pi}{8}$ & $\hat{\phi}^s_m=\frac{3.18\pi}{8}$ \\
\cline{3-5} & & s = 3 & $\hat{\phi}^s_m=\frac{3.24\pi}{8}$ & $\hat{\phi}^s_m=\frac{3.18\pi}{8}$ \\
\cline{2-5} & \multirow{2}{*}{256}& s = 2 &  $\hat{\phi}^s_m=\frac{2.85\pi}{8}$ & $\hat{\phi}^s_m=\frac{2.88\pi}{8}$ \\
\cline{3-5} & & s = 3 & $\hat{\phi}^s_m=\frac{2.85\pi}{8}$ & $\hat{\phi}^s_m=\frac{2.88\pi}{8}$ \\
\cline{2-5} \hline
\end{tabular} }}
\end{table}
Consider the system model (\ref{e7}) in which $M$ users
synchronously send their bits to the receiver through their
channels. It is assumed that each user's information consists of
codes of length $N$. It is also assumd that the signal to noise
ratio (SNR) is 0dB. In this example there is no power-unbalanced or
channel loss is assumed. The step-size of the NLMS algorithm in
modified LMS-PPIC method is $\mu=0.1(1-\sqrt{\frac{M-1}{M}})$ and
the set of step-sizes of the parallel NLMS algorithms in modified
PLMS-PPIC method are
$\Theta=\{0.01,0.05,0.1,0.2,\cdots,1\}(1-\sqrt{\frac{M-1}{M}})$,
i.e. $\mu_1=0.01(1-\sqrt{\frac{M-1}{M}}),\cdots,
\mu_4=0.2(1-\sqrt{\frac{M-1}{M}}),\cdots,
\mu_{12}=(1-\sqrt{\frac{M-1}{M}})$. Figure~\ref{Figexp1NonCoh}
illustrates the bit error rate (BER) for the case of two stages and
for $N=64$ and $N=256$. Simulations also show that there is no
remarkable difference between results in two stage and three stage
scenarios. Table~\ref{tabex5} compares the average channel phase
estimate of the first user in each stage and over $10$ runs of
modified LMS-PPIC and PLMS-PPIC, when the the number of users is
$M=15$.
\end{example}

Although LMS-PPIC and PLMS-PPIC, as well as their modified versions,
are structured based on the assumption of no near-far problem
(examples \ref{ex3} and \ref{ex4}), these methods and especially the
second one have remarkable performance in the cases of unbalanced
and/or time varying channels.

\begin{figure} \centering
\includegraphics[width=.45\textwidth]{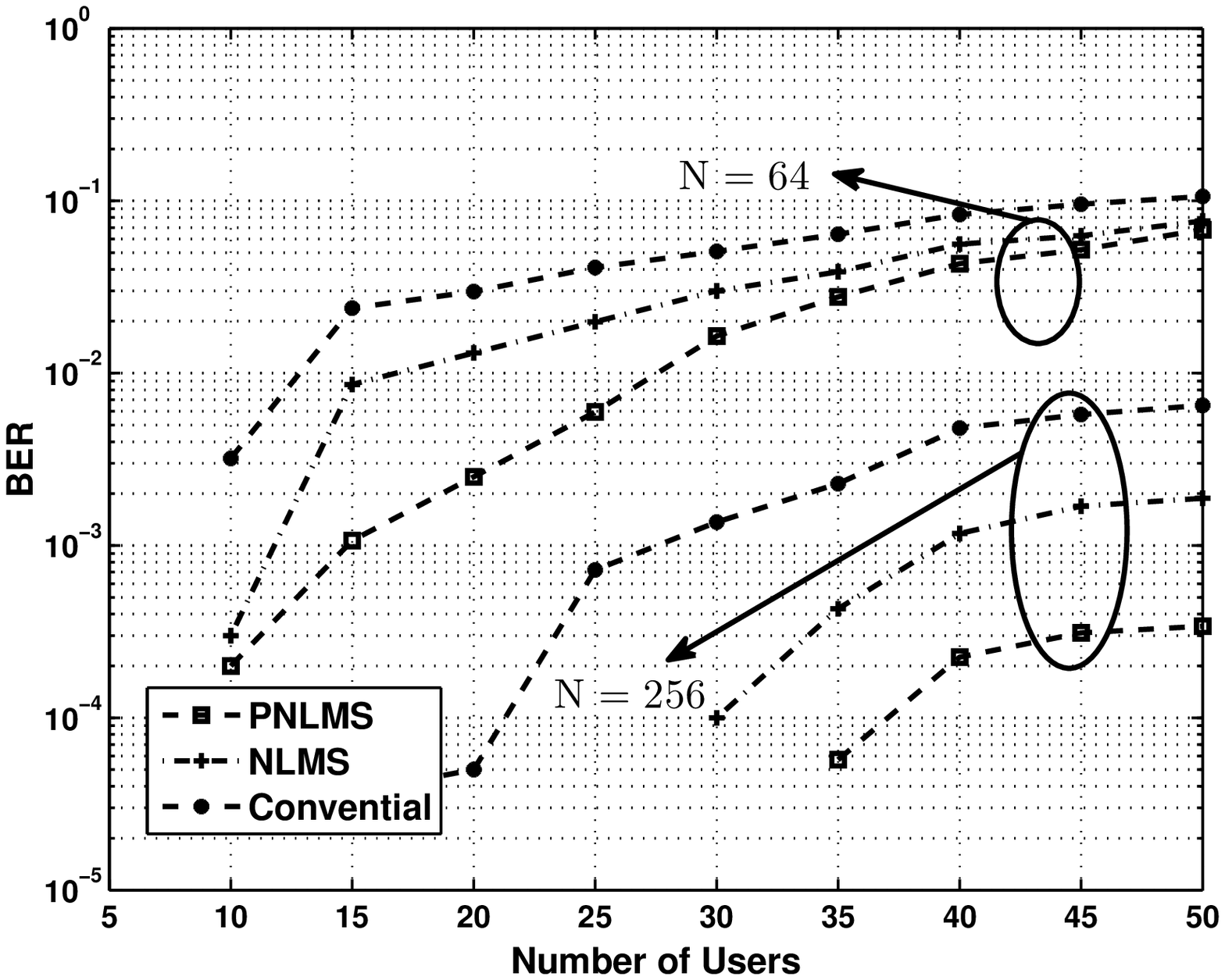}
\caption{The BER of the conventional, the modified LMS-PPIC and the
modified PLMS-PPIC methods versus the system load in unbalanced
channel, using two stages for $N=64$ and
$N=256$.}\label{Figexp2NonCoh}
\end{figure}

\begin{example}{\it Unbalanced channels}:
\label{ex3}
\begin{table}
\caption{Channel phase estimate of the first user (example
\ref{ex3})} \label{tabex6} \centerline{{
\begin{tabular}{|c|c|c|c|c|}
\hline
\multirow{6}{*}{\rotatebox{90}{$\phi_m=\frac{3\pi}{8},M=15~~$}} & N(Iteration) & Stage Number& NLMS & PNLMS  \\
&&&&\\
\cline{2-5} & \multirow{2}{*}{64}& s=2 &  $\hat{\phi}^s_m=\frac{2.45\pi}{8}$ & $\hat{\phi}^s_m=\frac{2.36\pi}{8}$ \\
\cline{3-5} & & s=3 & $\hat{\phi}^s_m=\frac{2.71\pi}{8}$ & $\hat{\phi}^s_m=\frac{2.80\pi}{8}$ \\
\cline{2-5} & \multirow{2}{*}{256}& s=2 &  $\hat{\phi}^s_m=\frac{3.09\pi}{8}$ & $\hat{\phi}^s_m=\frac{2.86\pi}{8}$ \\
\cline{3-5} & & s=3 & $\hat{\phi}^s_m=\frac{2.93\pi}{8}$ & $\hat{\phi}^s_m=\frac{3.01\pi}{8}$ \\
\cline{2-5} \hline
\end{tabular} }}
\end{table}
Consider example \ref{ex2} with power unbalanced and/or channel loss
in transmission system, i.e. the true model at stage $s$ is
\begin{equation}
\label{ve7} r(n)=\sum\limits_{m=1}^{M}\beta_m
w^s_m\alpha^{(s-1)}_m c_m(n)+v(n),
\end{equation}
where $0<\beta_m\leq 1$ for all $1\leq m \leq M$. Both the LMS-PPIC
and the PLMS-PPIC methods assume the model (\ref{e7}), and their
estimations are based on observations $\{r(n),X^s(n)\}$, instead of
$\{r(n),\mathbf{G}X^s(n)\}$, where the channel gain matrix is
$\mathbf{G}=\mbox{diag}(\beta_1,\beta_2,\cdots,\beta_m)$. In this
case we repeat example \ref{ex2}. We randomly get each element of
$G$ from $[0,0.3]$. Figure~\ref{Figexp2NonCoh} illustrates the BER
versus the number of users. Table~\ref{tabex6} compares the channel
phase estimate of the first user in each stage and over $10$ runs of
modified LMS-PPIC and modified PLMS-PPIC for $M=15$.
\end{example}

\begin{figure} \centering
\includegraphics[width=.45\textwidth]{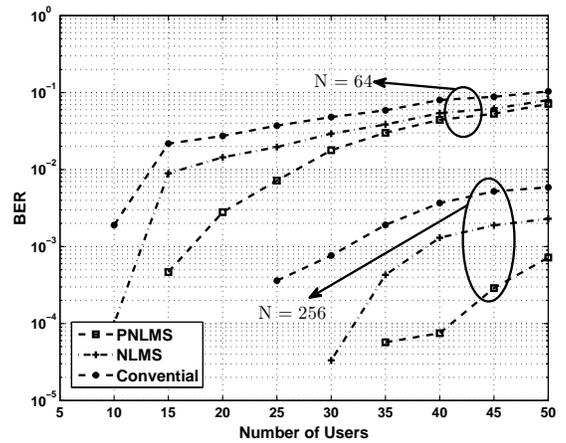}
\caption{The BER of the conventional, the modified LMS-PPIC and the
modified PLMS-PPIC methods versus the system load in time varying
Rayleigh fading channel, using two stages for $N=64$ and
$N=256$.}\label{Figexp3NonCoh}
\end{figure}

\begin{example}
\label{ex4} {\it Time varying channels}: Consider example \ref{ex2}
with time varying Rayleigh fading channels. In this case we assume
the maximum Doppler shift of $40$HZ, the three-tap
frequency-selective channel with delay vector of $\{2\times
10^{-6},2.5\times 10^{-6},3\times 10^{-6}\}$sec and gain vector of
$\{-5,-3,-10\}$dB. Figure~\ref{Figexp3NonCoh} shows the average BER
over all users versus $M$ and using two stages.
\end{example}

\section{Conclusion}\label{S6}

In this paper, parallel interference cancelation using adaptive
multistage structure and employing a set of NLMS algorithms with
different step-sizes is proposed, when just the quarter of the
channel phase of each user is known. In fact, the algorithm has been
proposed for coherent transmission with full information on channel
phases in \cite{cohpaper}. This paper is a modification on the
previously proposed algorithm. Simulation results show that the new
method has a remarkable performance for different scenarios
including Rayleigh fading channels even if the channel is
unbalanced.

\clearpage

\begin{table}
\caption{The procedure of the modified PLMS-PPIC method}
\label{tab4} \begin{center} {{\normalsize
\begin{tabular}{|l|l|l|l|l|l|}
\hline
\multirow{4}{*}{\rotatebox{90}{Initial Values~~~~~}} & {$\mbox{for}~~m=1,2,\cdots,M $} &\multicolumn{4}{|l|}{$\phi_m\in R_i,~~i=1,2,3,4 ~~~ \Longrightarrow$} \\
&&\multicolumn{4}{|l|}{} \\
&&\multicolumn{4}{|l|}{$~\hat{\phi}^0_m=\frac{(i-1)\pi+i\pi}{4}$} \\
&&\multicolumn{4}{|l|}{$~\alpha^{0}_m=\mbox{sign}\left\{\mbox{real}\left\{\sum\limits_{n=1}^{N} r(n)e^{-j\hat{\phi}^0_m}p_m(n)\right\}\right\}$} \\
&&\multicolumn{4}{|c|}{} \\
\hline
\multicolumn{2}{|l|}{$\mbox{for}~~s=1,2,\cdots,S $} &\multicolumn{4}{|l|}{$W^{s}(0)=[w^s_1(0),\cdots,w^s_M(0)]^T=[0,\cdots,0]^T$}\\
\cline{3-6} \multicolumn{2}{|l|}{}
&\multirow{4}{*}{\rotatebox{90}{PNLMS algorithm~~~~~~~~~~}}&{$\mbox{for}~~n=1,2,\cdots,N$}&\multicolumn{2}{|l|}{$X^{s}(n)=[\alpha^{(s-1)}_1c_1(n),\alpha^{(s-1)}_2c_2(n),\cdots,\alpha^{(s-1)}_Mc_M(n)]^T$}\\
\multicolumn{2}{|l|}{} &&&\multicolumn{2}{|l|}{$e(n)=r(n)-W^{s^T}(n-1)X^{s}(n)$} \\
\multicolumn{2}{|l|}{} &&&\multicolumn{2}{|l|}{$Z(n)=\frac{X^{s^*}(n)}{\|X^s(n)\|^2}e(n)$} \\
\multicolumn{2}{|l|}{} &&&\multicolumn{2}{|l|}{$\mbox{min}=\infty, l=1$} \\
\cline{5-6}
\multicolumn{2}{|l|}{} &&&$\mbox{for}~~k=1,2,\cdots,L$ &$W^{s}_k(n)=W^{s}(n-1)+\mu_{k}Z(n)$ \\
\multicolumn{2}{|l|}{} &&&&$\mbox{if}~~\sum\limits_{m=1}^{M}||w^s_{m,k}(n)|-1|<\mbox{min}:$ \\
\multicolumn{2}{|l|}{} &&&&$~~~~~\mbox{min}=\sum\limits_{m=1}^{M}||w^s_{m,k}(n)|-1|$\\
\multicolumn{2}{|l|}{} &&&&$~~~~~l=k$ \\
\cline{5-6}
\multicolumn{2}{|l|}{} &&&\multicolumn{2}{|l|}{$W^s(n)=W^s_l(n)$} \\
\cline{3-6} \multicolumn{2}{|l|}{}&\multirow{7}{*}{\rotatebox{90}{Phase Estimation~~~~}}&{$\mbox{for}~~m=1,2,\cdots,M$}&\multicolumn{2}{|l|}{$i=1,2,3,4 ~~~ \Longrightarrow$}\\
\multicolumn{2}{|l|}{}&&&\multicolumn{2}{|l|}{}\\
\multicolumn{2}{|l|}{}&&&\multicolumn{2}{|l|}{$~\hat{\phi}^s_m = \angle {w^s_m(N)}~~~~~~~~~\mbox{if}~ \angle{w^s_m(N)}, \phi_m\in R_i$}\\
\multicolumn{2}{|l|}{}&&&\multicolumn{2}{|l|}{$~\hat{\phi}^s_m = \angle {w^s_m(N)}+\pi~~~~ \mbox{if}~ \angle{w^s_m(N)}+\pi, \phi_m\in R_i$}\\
\multicolumn{2}{|l|}{}&&&\multicolumn{2}{|l|}{$~\hat{\phi}^s_m = \angle {w^s_m(N)}-\pi~~~~ \mbox{if}~ \angle{w^s_m(N)}-\pi, \phi_m\in R_i$}\\
\multicolumn{2}{|l|}{}&&&\multicolumn{2}{|l|}{$~\hat{\phi}^s_m = \frac{(i-1)\pi+i\pi}{4}~~~~~~~~~\mbox{if}~ \phi_m\in R_i,\angle{w^s_m(N)},\angle {w^s_m(N)}\pm\pi\notin R_i$}\\
\multicolumn{2}{|l|}{}&&&\multicolumn{2}{|l|}{}\\
%\multicolumn{2}{|l|}{}&&&\multicolumn{2}{|l|}{}\\
%\multicolumn{2}{|l|}{}&&&\multicolumn{2}{|l|}{}\\
\cline{3-6}
\multicolumn{2}{|l|}{}& \multicolumn{2}{|l|}{$\mbox{for}~~m=1,2,\cdots,M $} &\multicolumn{2}{|l|}{$q^{s}_{m}(n)=r(n)-\sum\limits_{m^{'}=1,m^{'}\ne m}^{M}w^{s}_{m^{'}}(N)\alpha^{(s-1)}_{m^{'}} p_{m^{'}}(n)$} \\
\multicolumn{2}{|l|}{}& \multicolumn{2}{|l|}{} &\multicolumn{2}{|l|}{$\alpha^{s}_m=\mbox{sign}\left\{\mbox{real}\left\{\sum\limits_{n=1}^{N}q^s_m(n)e^{-j\hat{\phi}^s_m}p_m(n)\right\}\right\}$} \\
\hline
\end{tabular} }}
\end{center}
\end{table}

\end{document}